\documentclass[aps,prb,showpacs,twocolumn,superscriptaddress]{revtex4-1}
\usepackage{amssymb}
\usepackage{graphicx}
\usepackage{appendix}
\usepackage{dcolumn}
\usepackage{bm}
\usepackage{amsmath}
\usepackage{epstopdf}
\usepackage{xcolor}
\usepackage{float}
\usepackage[english]{babel}
\usepackage[colorlinks,linkcolor=magenta,citecolor=blue,urlcolor=blue]{hyperref}

\begin{document}

\title{Real-complex transition driven by quasiperiodicity: a new universality class beyond $\mathcal{PT}$ symmetric one}
\author{Tong Liu}
\thanks{t6tong@njupt.edu.cn}
\affiliation{Department of Applied Physics, School of Science, Nanjing University of Posts and Telecommunications, Nanjing 210003, China}
\author{Xu Xia}
\thanks{1069485295@qq.com}
\affiliation{Chern Institute of Mathematics and LPMC, Nankai University, Tianjin 300071, China}

\date{\today}

\begin{abstract}
  We study a one-dimensional lattice model subject to non-Hermitian quasiperiodic potentials.
  Firstly, we strictly demonstrate that there exists an interesting dual mapping relation between $|a|<1$ and $|a|>1$ with regard to the potential tuning parameter $a$. The localization property of $|a|<1$ can be directly mapping to that of $|a|>1$, the analytical expression of the mobility edge of $|a|>1$ is therefore obtained through spectral properties of $|a|<1$. More impressive, we prove rigorously that even if the phase $\theta \neq 0$ in quasiperiodic potentials, the model becomes non-$\mathcal{PT}$ symmetric, however, there still exists a new type of real-complex transition driven by non-Hermitian disorder, which is a new universality class beyond $\mathcal{PT}$ symmetric class.

\end{abstract}

\maketitle

\section{Introduction}
In quantum physics, the conservation of energy and probability demands that a closed system exhibits real energies, which leads to all physical observables must be represented by Hermitian operators~\cite{Shankar}. However, when an open quantum system couples to a surrounding environment, the Hamiltonian of the system becomes non-Hermitian and the physical process can be effectively described through a complex eigenenergy~\cite{Lindblad}. Recently, non-Hermitian systems have been attracted growing interest motivated by experimental realization of photonic lattices~\cite{Photonic1,Photonic2,Photonic3,Photonic4} and ultracold atomic gases~\cite{ultracold1,ultracold2,ultracold3,ultracold4}. A celebrated paradigm in the non-Hermitian physics was parity-time ($\mathcal{PT}$) symmetric class, namely, a large class of non-Hermitian Hamiltonians, discovered by Bender and Boettcher~\cite{Bender1,Bender2}, can exhibit entirely real spectra as long as they commute with the $\mathcal{PT}$ operator. From an intuitive point of view, a necessary (but not sufficient) condition for $\mathcal{PT}$ symmetry to preserve is that the involved complex potential should satisfy $V(n)=V^*(-n)$ in the discrete lattice, where $n$ is the lattice site number~\cite{PT_1,PT_2,PT_3,PT_4,PT_5,PT_6}. Although the impact of $\mathcal{PT}$ symmetry in real quantum systems is still debated, broad research interests have been put into the study of $\mathcal{PT} $ symmetry-breaking transitions between real and complex eigenenergies in various non-Hermitian systems~\cite{Skin1,Skin2,Yao1,Yao2}.

Here a natural question arises: can other class of non-Hermitian systems host entirely real spectra beyond $\mathcal{PT}$ symmetric class? Disordered systems provide a potential candidate~\cite{random,quasiperiodic}. A paradigm to understand the Anderson localization in one dimension (1D) is the famous Aubry-Andr\'{e} (AA) model~\cite{AA,Harper}, one of the most studied example displaying a localization-delocalization transition~\cite{Laurent1,Laurent2,Laurent3,Laurent4}.
Another interesting aspect of AA-like models is the presence of mobility edges separating extended from localized states~\cite{mobility_1,mobility_2,mobility_3,Liu1}.
The combination of non-Hermitian properties and disordered systems is an interesting and active research topic~\cite{Longhi1,Longhi2,Gong,Kawabata}. A seminal work dealing with disorder in non-Hermitian systems is the Hatano-Nelson model~\cite{Hatano1,Hatano2}, in which
an asymmetric hopping caused by an imaginary gauge field results in a localization-delocalization transition~\cite{Hatano3}.
Other non-Hermitian models with either random or quasiperiodic disorder have been investigated, in which the localization properties of systems are the main concerns and contents~\cite{Jiang,LiuY,Zeng}.

With regard to the real-complex transition of spectra, in a pioneering study, Hamazaki et al. numerically found that a real-complex transition occurred in non-Hermitian interacting systems without $\mathcal{PT}$ symmetry~\cite{Hamazaki}, the real energy region was located in the localized phase of the model with time-reversal symmetry. In the further research, Liu et al. numerically demonstrated that the real energy region was located in the extended phase of a quasiperiodic $p$-wave superconductor chain without $\mathcal{PT}$ symmetry~\cite{Liu2}. However, these available studies are limited to numerical simulations, there are very few rigorous results to clarify the underlying physics of the real energy. In this work, we are devoted to introduce an exactly solvable model and analytically demonstrate that there exists a new type of real-complex transitions driven by non-Hermitian disorder, which is different from $\mathcal{PT}$ symmetric class.

\section{Model and The dual mapping}
We consider one-dimensional non-Hermitian quasiperiodic model, described by the following eigenvalue equation,
\begin{equation}
E \psi_n=\psi_{n+1}+\psi_{n-1}+\frac{V}{1-a e^{i (2\pi\alpha n+\theta)}} \psi_n,
\label{eq1}
\end{equation}
where $V$ is the complex potential strength, $\alpha$ is irrational, and $\psi_n$ is the amplitude of wave function at the $n$th lattice. We choose to unitize the nearest-neighbor hopping amplitude and a typical choice for parameter $\alpha$ is $\alpha=(\sqrt{5}-1)/2$. The phase $\theta$ locates in the center of this study, which is not simply taken as zero. $|a|\neq1$ is the potential tuning parameter.
When $|a|<1$ and $\theta=0$, Eq.~(\ref{eq1}) is $\mathcal{PT}$ symmetric, Liu et al.~\cite{Liu1} derived the analytical expression of the energy-dependent mobility edge by means of a generalized self-duality relation. In this work we demonstrate that the $|a|>1$ case is the dual counterpart of the $|a|<1$ case, and the localization property of $|a|<1$ can be directly mapping to that of $|a|>1$. Moreover, we prove rigorously that when $\theta \neq 0$, Eq.~(\ref{eq1}) becomes non-$\mathcal{PT}$ symmetric, however, there still exists a real-complex transition in the energy spectrum, which is unconventional and different from $\mathcal{PT}$ symmetric class.

Now we unveil a dual mapping relation between $|a|<1$ and $|a|>1$ of Eq.~(\ref{eq1}).
Note the non-Hermitian quasiperiodic potential $\frac{V}{1-ae^{i x}}$ can be written as Taylor's expansion,
\begin{equation}
\frac{V}{1-a e^{i x}}=V \sum_{m=0}^{\infty} a^{m} e^{i m x},
\label{eq2}
\end{equation}
then,
\begin{equation}
\frac{V}{1-a e^{i x}}=\frac{-V}{1-1/a e^{-i x}}+V,
\label{eq3}
\end{equation}
thus, spectral properties of $\frac{V}{1-ae^{i x}}$ ($|a|<1$) can be dual mapping to these of the potential $\frac{-V}{1-ae^{-i x}}+V$ ($|a|>1$).
Consequently, utilizing the mobility edge energy $E_{c}=a+1/a$ ($|a|<1$), with regard to $|a|>1$ we can directly obtain the exact explicit form of the mobility edge:
\begin{equation}
\label{eq4}
E_m=V-a-1/a.
\end{equation}

\begin{figure}
  \centering
  \includegraphics[width=0.5\textwidth]{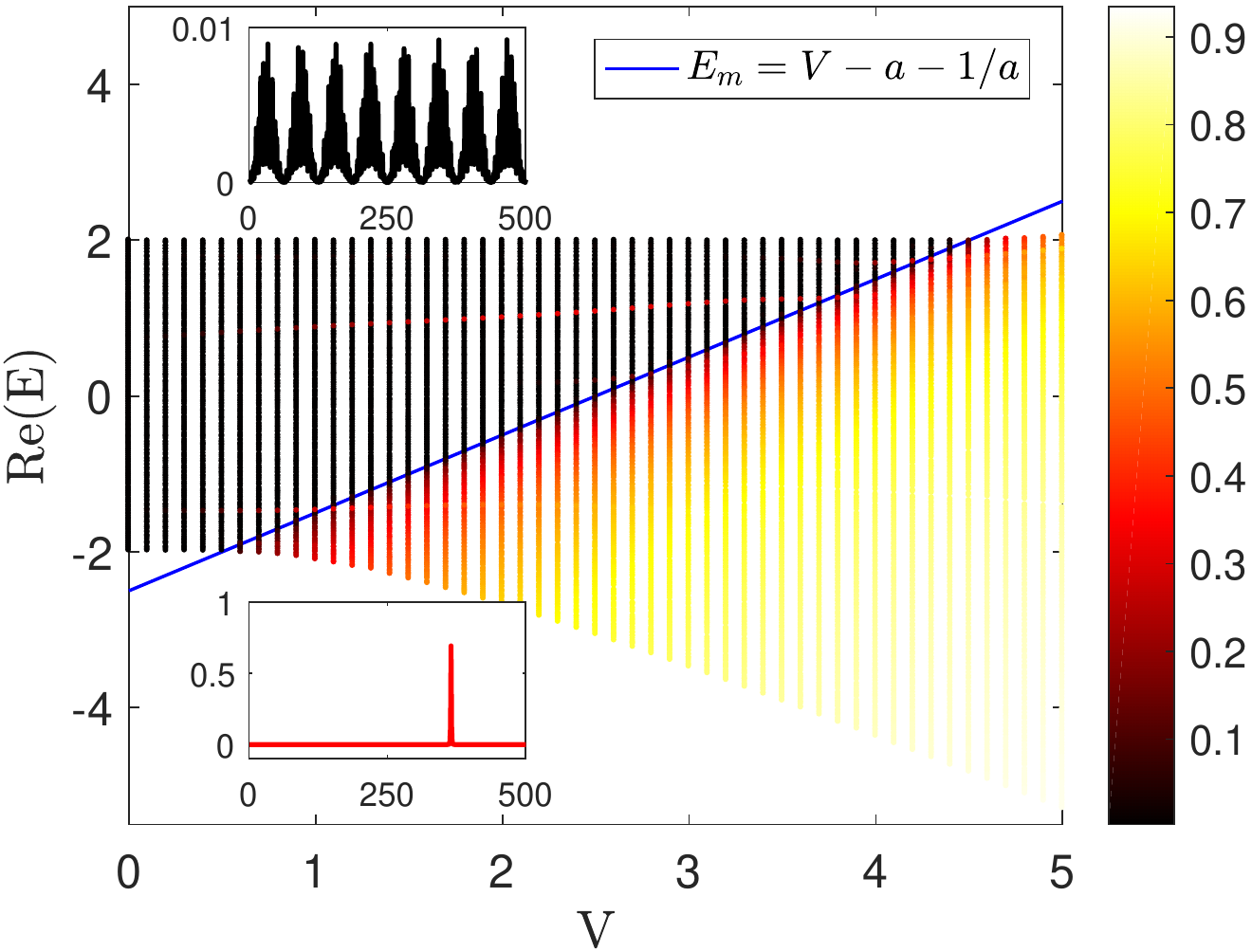}\\
  \caption{(Color online) The real part of eigenvalues of Eq.~(\ref{eq1}) and IPR as a function of $V$ with the parameter $a=2$. The total number of sites is set to be $L=500$. The blue solid lines represent the boundary between spatially localized and delocalized states, i.e., the mobility edge $E_m=V-a-1/a$. In the inset, we plot the typical probability density $|\psi|$ with eigenvalues above and below $E_m$ ($V=2$), as expected, they are the extended state (black) and the localized state (red), respectively.}
  \label{fig1}
\end{figure}

To support the analytical result given above, we now present detailed numerical analysis of Eq.~(\ref{eq1}). In the disordered system, the localization property of wave functions can be measured by the inverse participation ratio (IPR)~\cite{IPR}. For any given normalized wave function, the corresponding IPR is defined as $\text{IPR} =\sum_{n=1}^{L} \left|\psi_{n}\right|^{4},$
which measures the inverse of the number of sites being occupied by particles. It is well known that the IPR of an extended state scales
like $L^{-1}$ which approaches zero in the thermodynamic limit. However, for a localized state, since only finite number of sites are
occupied, the IPR is finite even in the thermodynamic limit. In Fig.~\ref{fig1} we show the numerical IPR diagram in the $[{\rm Re}(E),V]$ plane, different colours of the eigenvalue curves indicate different magnitudes of the IPR of the corresponding wave functions. The black eigenvalue curves denote the extended states, and the bright yellow eigenvalue curves denote the localized states. It is clearly demonstrating a mobility edge separating localized from extended states along the blue line defined by Eq.~(\ref{eq4}).

\section{Real-complex transition}
Besides the localization, the most novel discovery of this work is that there exists a new type of real-complex transitions in Eq.~(\ref{eq1}) driven by non-Hermitian disorder, rather than $\mathcal{PT}$ symmetry.

For the sake of simplicity, here we focus on the mathematical proof of the $|a|<1$ case. First let us multiply both sides of Eq.~(\ref{eq1}) by $e^{i  n k}$ and sum over $n$. After setting $f(k)=\sum_n e^{i  n k}\psi_n$, one obtains
\begin{equation}\label{eq5}
a[E-2\cos(k+2\pi\alpha)] f(k+2\pi\alpha)=[E-V-2\cos(k)]f(k),
\end{equation}

According to Sarnak's method~\cite{Sarnak}, the spectrum of Eq.~(\ref{eq5}) is govern by a charateristic function defined as
\begin{equation}
\begin{aligned}
G(E)&=\frac{1}{2 \pi} \int_{0}^{2 \pi} \log \left|E-V-2\cos(k)\right| d k\\
&-\frac{1}{2 \pi} \int_{0}^{2 \pi} \log \left|E-2\cos(k+2\pi\alpha)\right| d k,
\label{eq6}
\end{aligned}
\end{equation}

Here, we don't show more complicated mathematical proofs, but directly quote Sarnak's Lemma~\cite{Sarnak}:

(i) There is no energy spectrum when $G(E)<\log |a|$.

(ii) When $G(E)=\log |a|$, Eq.~(\ref{eq5}) has dense spectrum, the corresponding wave functions are localized.

(iii) When $G(E)>\log |a|$, wave functions of Eq.~(\ref{eq5}) are extended. However, the spectrum of the system must be within the energy interval $U_{E}=[V-2,V+2]$, which is derived by the equation $E-V-2\cos(k)=0$.

The conclusion (iii) is the central result, it demonstrates that when the system is in the extended phase, the spectrum must be restricted to a real interval, not a complex number, it should be noted that the above proof does not involve $\mathcal{PT}$ symmetry.

When the spectrum $E$ is restricted in $U_{E}$, Eq.~(\ref{eq6}) can be reduced to
\begin{equation}\label{eq7}
G(E)=\log|\frac{E-\sqrt{E^2-4}}{2}|
\end{equation}

From the conclusion (ii), we can obtain the metal-insulator transition boundary, i.e., the mobility edge $E_{c}=a+1/a$ derived by the equation $G(E_c)=\ln(a)$. Remarkably, this expression is exactly alike that derived by the generalized self-duality relation. And the total real spectrum region is enclosed by three lines $E_{min}=V-2$, $E_{max}=V+2$ and $E_{c}=a+1/a$.
\begin{figure}[t]
	\centering
	\includegraphics[width=0.48\textwidth]{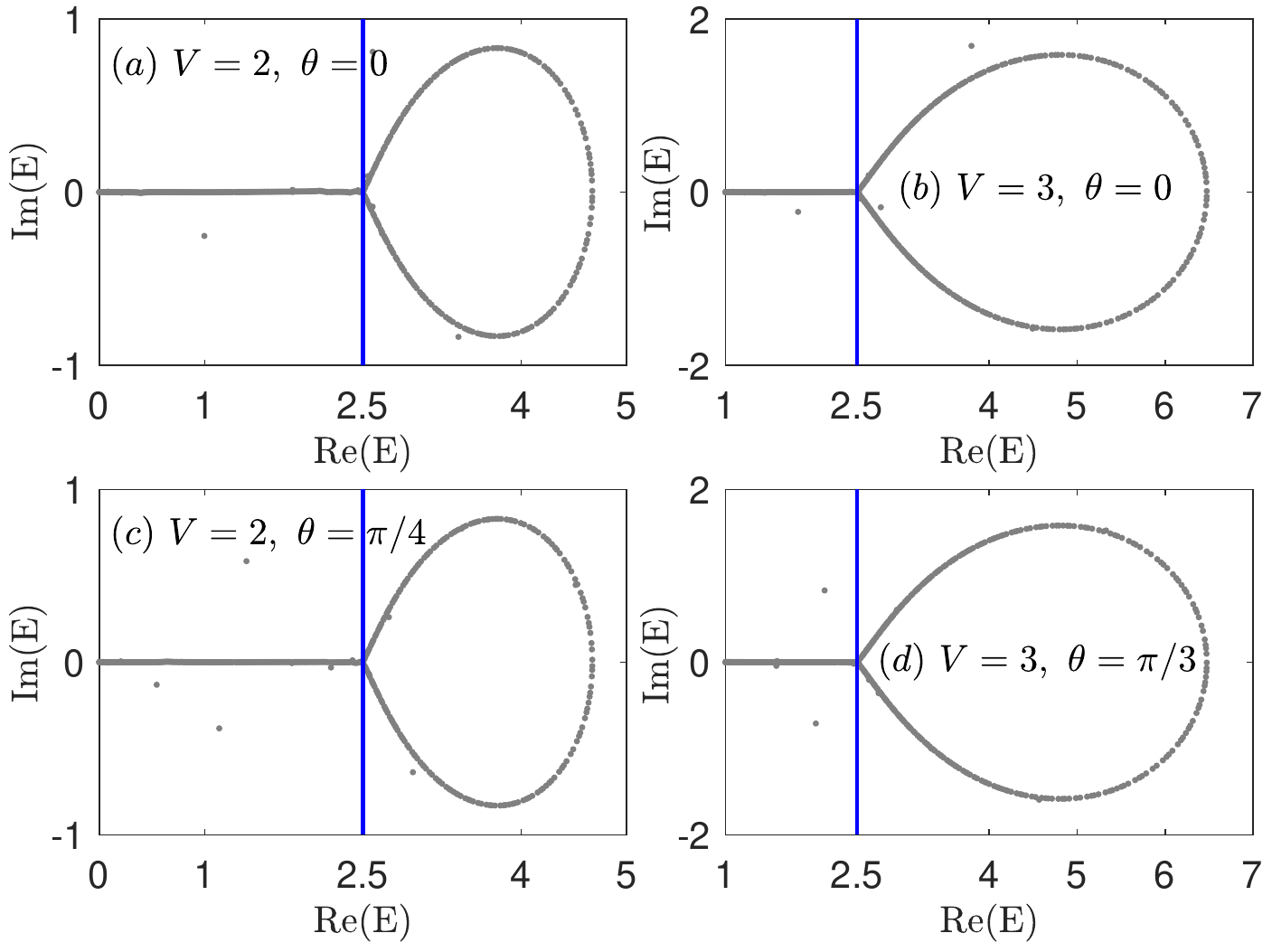}\\
	\caption{(Color online) The eigenvalues of Eq.~(\ref{eq1}) with the parameter $a=0.5$ under open boundry conditions. The total number of sites is set to be $L=500$. (a) and (c) use the same potential strength $V=2$ but different phase $\theta$ for comparison. Similarly, (b) and (d) use the same potential strength $V=3$ but different phase $\theta$ for comparison. The blue solid lines represent the boundary between the real and complex energy spectrum, it is clearly shown that the real spectrum region of four cases is restricted in $E_{min}=V-2$ and $E_{c}=a+1/a$.}
	\label{fig2}
\end{figure}

The above derivation ignores the phase $\theta$. When $\theta =0$, Eq.~(\ref{eq1}) still preserve the $\mathcal{PT}$ symmetry, this may be the reason why most works~\cite{LiuY,Zeng,PTPT1,PTPT2} classified this type of real-complex transition as the $\mathcal{PT}$ symmetry. To clarify the difference, we further focus on the $\theta \neq 0$ case, now Eq.~(\ref{eq1}) obviously does not satisfy the $\mathcal{PT}$ symmetry. Concerning the phase $\theta$ in quasiperiodic potentials, Simon uses the mathematical ergodic theory to prove that the eigenvalues are almost independent of the phase~\cite{Simon}. At the same time, he also pointed out that the unique ergodic is completely independent of the phase, but he did not give the details of the proof. We add here the proof that eigenvalues of our model is completely phase-independent.
We mark the tridiagonal matrix of the Hamiltonian as $H_{\theta_{1}}$ and $H_{\theta_{2}}$ when two phases are $\theta_{1}$ and $\theta_{2}$, respectively.
All eigenvalues of $H_{\theta_{1}}$ and $H_{\theta_{2}}$ are denoted as $\Sigma_1$ and $\Sigma_2$, consequently.

Then for any integer $n$, there exists the following estimate
\begin{equation}
\begin{aligned}
\|  H_{\theta_{1}}-H_{\theta_{2}} \| &\leqslant   \frac{2V}{(1-a)^2} |\sin (\theta_{1}-\theta_{2}-n w){\pi}|\\
&\leqslant \frac{2\pi V}{(1-a)^2} {\left|\theta_{1}-\theta_{2}-n \omega\right|_{\mathbb{R}/\mathbb{Z}}},
\end{aligned}
\end{equation}
thus, we obtain
\begin{equation}
h(\Sigma_1,\Sigma_2)\leq\frac{2\pi V}{(1-a)^2} {\left|\theta_{1}-\theta_{2}-n \omega\right|_{\mathbb{R}/\mathbb{Z}}}.
\end{equation}
where $h$ represents the minimum distance of two sets $\Sigma_1$ and $\Sigma_2$.
Because the frequency $\alpha$ is an irrational number, we can take a list of $n_k$ to make $ \left|\theta_{1}-\theta_{2}-n_k \omega\right|_{\mathbb{R}/\mathbb{Z}}$ approach zero, which means $h=0$. Consequently, we strictly prove $\Sigma_1=\Sigma_2$, irrelevant to $\theta_{1}$ and $\theta_{2}$. Thus even if $\theta \neq 0$, i.e., it doesn't exist the $\mathcal{PT}$ symmetry, the real spectrum of the system still stays the same, these results strongly demonstrate that the real-complex transition driven by $\exp(i x)$-like non-Hermitian disorder belongs to a unique class, rather than the $\mathcal{PT}$ symmetric class.
\begin{figure}[t]
	\centering
	\includegraphics[width=0.48\textwidth]{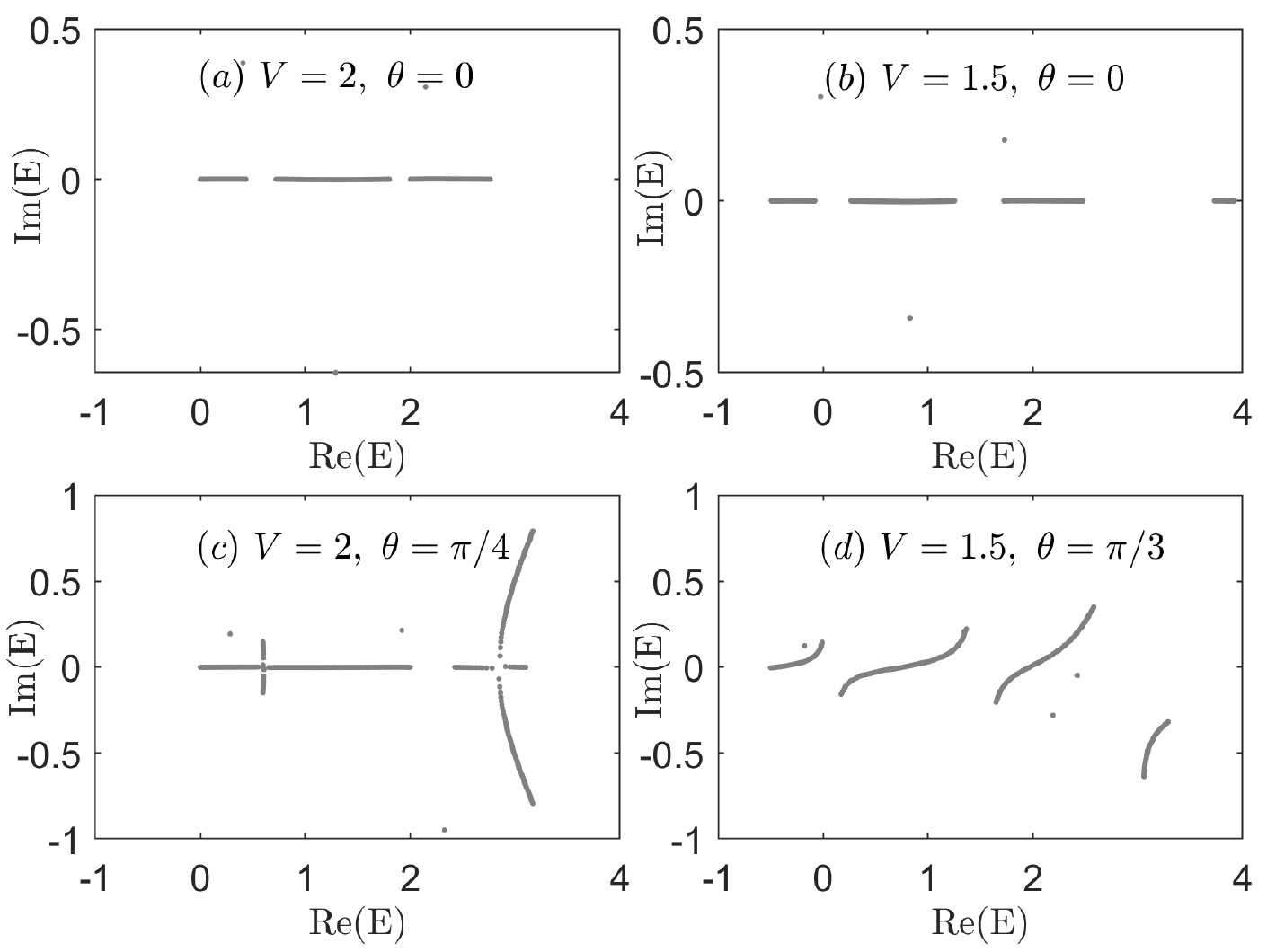}\\
	\caption{(Color online) The eigenvalues of Eq.~(\ref{eq1}) with the parameter $a=0.5$ under open boundry conditions, $\alpha$ is set to a rational number $1/4$. The total number of sites is set to be $L=500$. In this situation, the $\theta \neq 0$ case has different spectrum with the $\theta = 0$ case, and the real-complex transition is within the logical framework of the $\mathcal{PT}$ symmetry broken.}
	\label{fig3}
\end{figure}

In order to verify the theoretical analysis, we implement numerical calculations of eigenvalues. As shown in Fig.~\ref{fig2}, even if the phase $\theta$ is different, the eigenvalues of the system with the same $V$ are same, and the mobility edge $E_{c}=a+1/a$ exactly separates the real from complex spectrum, thus numerical results are in excellent agreement with the analytical results. We have also checked other combinations of parameters and get the same results as expected. In addition, there exist several deviated points in Fig.~\ref{fig2}, this is because, not like finite periodic systems, disordered systems are statistical and based on large samples. The proportion of deviated points will decrease sharply and be almost negligible with the increase of sample size.

A possible reason of being classified as the $\mathcal{PT}$ symmetric class in previous works is to ignore the difference of the rational and irrational of the parameter $\alpha$. For $\alpha$ being rational, the system is perfectly periodic, in this situation the real-complex transition is indeed driven by the $\mathcal{PT}$ symmetry, see Fig.~\ref{fig3}. When the phase $\theta=0$, the system preserves the $\mathcal{PT}$ symmetry, the eigenvalues are real, while the phase $\theta \neq 0$, the system doesn't preserve the $\mathcal{PT}$ symmetry, the eigenvalues are complex. For $\alpha$ being irrational, as we proved above, this real-complex transition violates the logical framework of the $\mathcal{PT}$ symmetry broken, hence can't be classified as the $\mathcal{PT}$ symmetric class.

\section{Summary}
In summary, we have shown an interesting dual mapping between $|a|<1$ and $|a|>1$ in a non-Hermitian quasiperiodic model, the spectral properties of the two can be mapping to each other. More impressively, we mathematically prove that the eigenenergies in the extended phase are fully real, and an unconventional real-complex transition occurs accompanied by the metal-insulator transition. The meaning of unconventionality is that even if our model doesn't preserve the conventional $\mathcal{PT}$ symmetry, the real-complex transition in the spectrum also appears exactly as our theoretical predictions.
Our findings act in cooperation with the previous numerical results of non-Hermitian disorder, and establish that the real-complex transition driven by non-Hermitian disorder is a new universality class, which is different from $\mathcal{PT}$ symmetric class.

\begin{acknowledgments}
T. L.~acknowledges the Natural Science Foundation of Jiangsu Province (Grant No.~BK20200737) and NUPTSF (Grant No.~NY220090 and No.~NY220208). X. X.~is supported by Nankai Zhide Foundation.
\end{acknowledgments}






\begin{references}
\bibitem{Shankar} R. Shankar, Principles of Quantum Mechanics (Plenum Press, 1994).
\bibitem{Lindblad} G. Lindblad, On the generators of quantum dynamical semigroups, Commun. Math. Phys. \textbf{48}, 119 (1976).

\bibitem{Photonic1} L. Feng, R. El-Ganainy, and L. Ge, Non-Hermitian photonics based on parity time symmetry, Nat. Photonics \textbf{11}, 752 (2017).
\bibitem{Photonic2} A. Cerjan, S. Huang, M. Wang, K. P. Chen, Y. Chong, and M. C. Rechtsman, Experimental realization of a Weyl exceptional ring, Nat. Photonics \textbf{13}, 623 (2019).
\bibitem{Photonic3} L. Xiao, X. Zhan, Z. H. Bian, K. K. Wang, X. Zhang, X. P.Wang, J. Li, K. Mochizuki, D. Kim, N. Kawakami, W. Yi, H. Obuse, B. C. Sanders, and P. Xue, Observation of topological edge states in parity time symmetric quantum walks, Nat. Phys. \textbf{13}, 1117 (2017).
\bibitem{Photonic4} J. M. Zeuner, M. C. Rechtsman, Y. Plotnik, Y. Lumer, S. Nolte, M. S. Rudner, M. Segev, and A. Szameit, Observation of a Topological Transition in the Bulk of a Non-Hermitian System, Phys. Rev. Lett. \textbf{115}, 040402 (2015).
\bibitem{ultracold1} I. Bloch, J. Dalibard, and W. Zwerger, Many-body physics with ultracold gases, Rev. Mod. Phys. \textbf{80}, 885 (2008).
\bibitem{ultracold2} J. Li, A. K. Harter, J. Liu, L. de Melo, Y. N. Joglekar, and L. Luo, Observation of parity-time symmetry breaking transitions in a dissipative Floquet system of ultracold atoms, Nat. Commun. \textbf{10}, 855 (2019).
\bibitem{ultracold3} W. Gou, T. Chen, D. Xie, T. Xiao, T.-S. Deng, B. Gadway, W. Yi, and B. Yan, Tunable Nonreciprocal Quantum Transport through a Dissipative Aharonov-Bohm Ring in Ultracold Atoms, Phys. Rev. Lett. \textbf{124}, 070402 (2020).
\bibitem{ultracold4} F. Alex An, E. J. Meier, and B. Gadway, Engineering a Flux-Dependent Mobility Edge in Disordered Zigzag Chains, Phys. Rev. X \textbf{8}, 031045 (2018).

\bibitem{Bender1} C. M. Bender and S. Boettcher, Real Spectra in Non-Hermitian Hamiltonians Having PT Symmetry, Phys. Rev. Lett. \textbf{80}, 5243 (1998).
\bibitem{Bender2} C. M. Bender, Making sense of non-Hermitian Hamiltonians, Rep. Prog. Phys. \textbf{70}, 947 (2007).

\bibitem{PT_1} C. Yuce, Topological phase in a non-Hermitian PT symmetric system, Phys. Lett. A \textbf{379}, 1213 (2015).
\bibitem{PT_2} D.-W. Zhang, L.-Z. Tang, L.-J. Lang, H. Yan, and S.-L. Zhu, Non-Hermitian topological Anderson insulators, Sci. China-Phys. Mech. Astron. \textbf{63}, 267062 (2020).
\bibitem{PT_3} D. S. Borgnia, A. J. Kruchkov, and R.-J. Slager, Non-Hermitian Boundary Modes and Topology, Phys. Rev. Lett. \textbf{124}, 056802 (2020).
\bibitem{PT_4}  K. Yamamoto, M. Nakagawa, K. Adachi, K. Takasan, M. Ueda, and N. Kawakami, Theory of Non-Hermitian Fermionic Superfluidity with a Complex-Valued Interaction, Phys. Rev. Lett. \textbf{123}, 123601 (2019).
\bibitem{PT_5}  R. El-Ganainy, K. G. Makris, M. Khajavikhan, Z. H. Musslimani, S. Rotter, and D. N. Christodoulides, Non-Hermitian physics and PT symmetry, Nat. Phys. \textbf{14}, 11 (2018).
\bibitem{PT_6}  H. Shen, B. Zhen, and L. Fu, Topological Band Theory for Non-Hermitian Hamiltonians, Phys. Rev. Lett. \textbf{120}, 146402 (2018).

\bibitem{Skin1} N. Okuma, K. Kawabata, K. Shiozaki, and M. Sato, Topological Origin of Non-Hermitian Skin Effects, Phys. Rev. Lett. \textbf{124}, 086801 (2020).
\bibitem{Skin2} K. Zhang, Z. Yang, and C. Fang, Correspondence between Winding Numbers and Skin Modes in Non-Hermitian Systems, Phys. Rev. Lett. \textbf{125}, 126402 (2020).
\bibitem{Yao1} S. Yao and Z. Wang, Edge states and topological invariants of non-Hermitian systems, Phys. Rev. Lett. \textbf{121}, 086803 (2018).
\bibitem{Yao2} S. Yao, F. Song, and Z. Wang, Non-Hermitian chern bands, Phys. Rev. Lett. \textbf{121}, 136802 (2018).

\bibitem{random}J. Billy, V. Josse, Z. Zuo, A. Bernard, B. Hambrecht, P. Lugan, D. Cl\'{e}ment, L. Sanchez-Palencia, P. Bouyer, and A. Aspect,
Direct observation of Anderson localization of matter waves in a controlled disorder, Nature (London) \textbf{453}, 891 (2008).
\bibitem{quasiperiodic}G. Roati, C. D. Errico, L. Fallani, M. Fattori, C. Fort, M. Zaccanti, G. Modugno, M. Modugno, and M. Inguscio, Nature (London) \textbf{453}, 895 (2008).

\bibitem{AA} S. Aubry and G. Andr\'{e}, Analyticity breaking and Anderson localization in incommensurate lattices, Ann. Isr. Phys. Soc. \textbf{3}, 133 (1980).
\bibitem{Harper} P. G. Harper, The General Motion of Conduction Electrons in a Uniform Magnetic Field, with Application to the Diamagnetism of Metals, Proc. Phys. Soc. London Sect. A \textbf{68}, 874 (1955).

\bibitem{Laurent1} F. Jendrzejewski, A. Bernard, K. M\"{u}ller, P. Cheinet, V.Josse, M. Piraud, L. Pezz\'e, L. Sanchez-Palencia, A. Aspect, and P. Bouyer, Three-dimensional localization of ultracold atoms in an optical disordered potential, Nat. Phys. \textbf{8}, 398 (2012).
 \bibitem{Laurent2} H. Yao, H. Khouldi, L. Bresque, and L. Sanchez-Palencia, Critical behavior and fractality in shallow one-dimensional quasiperiodic potentials, Phys. Rev. Lett. \textbf{123}, 070405 (2019). %
\bibitem{Laurent3} H. Yao, T. Giamarchi, and L. Sanchez-Palencia, Lieb-Liniger bosons in a shallow quasiperiodic potential: Bose glass phase and fractal Mott Lobes, Phys. Rev. Lett. \textbf{125}, 060401 (2020). %
\bibitem{Laurent4} L. Sanchez-Palencia, Ultracold gases: At the edge of mobility, Nat. Phys. \textbf{11}, 525 (2015).%

\bibitem{mobility_1} J. Biddle and S. Das. Sarma, Predicted mobility edges in one-dimensional incommensurate optical lattices: An exactly solvable model of Anderson localization, Phys. Rev. Lett. \textbf{104}, 070601 (2010).
\bibitem{mobility_2} S. Ganeshan, J. H. Pixley, and S. Das Sarma, Nearest neighbor tight binding models with an exact mobility edge in one dimension, Phys. Rev. Lett. {\bf 114}, 146601 (2015).
\bibitem{mobility_3} Y. Wang, X. Xia, L. Zhang, H. Yao, S. Chen, J. You, Q. Zhou, and X.-J. Liu, One dimensional quasiperiodic mosaic lattice with exact mobility edges, Phys. Rev. Lett. \textbf{125}, 196604 (2020).%
\bibitem{Liu1} T. Liu, H. Guo, Y. Pu, and S. Longhi, Generalized Aubry-Andr\'{e} self-duality and Mobility edges in non-Hermitian quasi-periodic lattices, Phys. Rev. B \textbf{102}, 024205 (2020).

\bibitem{Longhi1} S. Longhi, Topological phase transition in non-Hermitian quasicrystals, Phys. Rev. Lett. {\bf 122}, 237601 (2019).
\bibitem{Longhi2} S. Longhi, Metal-insulator phase transition in a non-Hermitian Aubry-Andr\'{e}-Harper Model, Phys. Rev. B {\bf 100}, 125157 (2019).
\bibitem{Gong} Z. Gong, Y. Ashida, K. Kawabata, K. Takasan, S. Higashikawa, and M. Ueda, Topological Phases of Non-Hermitian Systems, Phys. Rev. X \textbf{8}, 031079 (2018).
\bibitem{Kawabata} K. Kawabata, K. Shiozaki, M. Ueda, and M. Sato, Symmetry and topology in non-Hermitian physics, Phys. Rev. X \textbf{9}, 041015 (2019).

\bibitem{Hatano1} N. Hatano and D. R. Nelson,  Localization Transitions in Non-Hermitian Quantum Mechanics, Phys. Rev. Lett. {\bf 77}, 570 (1996).
\bibitem{Hatano2} N. Hatano and D. R. Nelson, Vortex pinning and non-Hermitian quantum mechanics, Phys. Rev. B {\bf 56}, 8651 (1997).
\bibitem{Hatano3} N. Hatano and D. R. Nelson, Non-Hermitian Delocalization and Eigenfunctions, Phys. Rev. B {\bf 58}, 8384 (1998).

\bibitem{Jiang} H. Jiang, L.-J. Lang, C. Yang, S.-L. Zhu, and S. Chen, Interplay of non-Hermitian skin effects and Anderson localization in non-reciprocal quasi-periodic lattices, Phys. Rev. B {\bf 100}, 054301 (2019).
\bibitem{LiuY} Y. Liu, X.-P. Jiang, J. Cao, and S. Chen, Non-Hermitian mobility edges in one-dimensional quasicrystals with parity-time symmetry, Phys. Rev. B {\bf 101}, 174205 (2020).
\bibitem{Zeng} Q.-B. Zeng, Y. B. Yang, and Y. Xu, Topological phases in non-Hermitian Aubry-Andr\'{e}-Harper models, Phys. Rev. B {\bf 101}, 020201 (2020).

\bibitem{Hamazaki} R. Hamazaki, K. Kawabata, and M. Ueda, Non-Hermitian Many-Body Localization, Phys. Rev. Lett. \textbf{123}, 090603 (2019).
\bibitem{Liu2}  T. Liu, S. Cheng, H. Guo, and G. Xianlong, Fate of Majorana zero modes, exact location of critical states, and unconventional real-complex transition in non-Hermitian quasiperiodic lattices, Phys. Rev. B \textbf{103}, 104203 (2021).

\bibitem{IPR} D. J. Thouless, A relation between the density of states and range of localization for one dimensional random systems, J. Phys. C: Solid State Phys. {\bf 5}, 77 (1973).

\bibitem{Sarnak} P. Sarnak, Spectral behavior of quasi periodic potentials, Commun. Math. Phys. {\bf 84}, 377 (1982).

\bibitem{PTPT1} S. Longhi, Phase transitions in a non-Hermitian Aubry-Andr\'e-Harper model, Phys. Rev. B {\bf 103} 054203 (2021).
\bibitem{PTPT2} Y. Liu,  Y. Wang,  X.-J. Liu,  Q. Zhou, and S. Chen, Exact mobility edges, PT-symmetry breaking and skin effect in one-dimensional non-Hermitian  quasicrystals, Phys. Rev. B {\bf 103}, 014203 (2021).

\bibitem{Simon} H. L. Cycon, R. G. Froese, W. Kirsch, and B. Simon, Schr\"odinger operators: With application to quantum mechanics and global geometry[M]. Springer, 2009.

\bibitem{RE1} A. Schreiber, K. N. Cassemiro, V. Potocek, A. Gabris, I. Jex, and Ch. Silberhorn, Decoherence and disorder in quantumwalks: From ballistic spread to localization, Phys. Rev. Lett. {\bf 106}, 180403 (2011).
\bibitem{RE2} I. D. Vatnik, A. Tikan, G. Onishchukov, D. V. Churkin, and A. A. Sukhorukov, Anderson localization in synthetic photonic lattices, Sci. Rep. {\bf 7}, 4301 (2017).
\bibitem{Derevyanko} S. Derevyanko, Disorder-aided pulse stabilization in dissipative synthetic photonic lattices, Sci. Rep. {\bf 9}, 12883 (2019).

\end{references}
\end{document}